\documentclass[usenatbib]{mn2e}
\usepackage{graphicx,fixltx2e}
\usepackage{wasysym}
\usepackage{color}

\def\rpd{\hbox{rad\,d$^{-1}$}}

\def\chisqr{\hbox{$\chi^2_{\rm r}$}}
\def\mSun{\hbox{${\rm M}_{\odot}$}}

\def\sn{\hbox{S/N}}
\def\vrad{\hbox{$v_{\rm rad}$}}
\def\ms{\hbox{m\,s$^{-1}$}}

\def\kms{\hbox{km\,s$^{-1}$}}
\def\vsini{\hbox{$v \sin i$}}

\def\ptt{\hbox{$10^{-4} I_{\rm c}$}}

\def\degr{\hbox{$^\circ$}}

\def\omeq{\hbox{$\Omega_{\rm eq}$}}
\def\dom{\hbox{$d\Omega$}}

\newcommand{\caii}{\hbox{Ca$\;${\sc ii}}~H~\&\ K~}

\newcommand{\halpha}{$\rm H\alpha$~}
\begin{document}

\title[]
{Magnetic field, differential rotation and activity of the hot-Jupiter hosting star HD~179949
 }

\makeatletter

\def\newauthor{%
  \end{author@tabular}\par
  \begin{author@tabular}[t]{@{}l@{}}}
\makeatother
\author[R.~Fares et al.]{\vspace{1.7mm} 
R.~Fares$^{1,2,3}$\thanks{E-mail:
rim.fares@ast.obs-mip.fr},
J.-F.~Donati$^2$, C.~Moutou$^3$,  M.~Jardine$^1$, A.C.~Cameron$^1$, A.F.~Lanza$^4$\\ 
\vspace{1.7mm}
{\hspace{-1.5mm}\LARGE\rm
D.~Bohlender$^5$, S.~Dieters$^6$, A.F.~Mart\'{i}nez~Fiorenzano$^7$, A.~Maggio$^8$, I.~Pagano$^4$}\\
{\hspace{-1.5mm}\LARGE\rm E.L.~Shkolnik$^9$  } \\
$^1$ School of Physics and Astronomy, Univ.\ of St~Andrews, St~Andrews, Scotland KY16 9SS, UK \\
$^2$ LATT--UMR 5572, CNRS \& Univ.\ P.~Sabatier, 14 Av.\ E.~Belin, F--31400 Toulouse, France \\
$^3$ LAM--UMR 6110, CNRS \& Univ.\ de Provence, 38 rue Fr\'ederic Juliot-Curie, F--13013 Marseille, France \\
$^4$ INAF-Osservatorio Astrofisico di Catania, via S. Sofia, 78 - 95123 Catania. Italy\\
$^5$ HIA/NRC, 5071 West Saanich Road, Victoria, BC V9E 2E7, Canada \\ 
$^6$ School of Mathematics and Physics, University of Tasmania, PB 37 GP0 Hobart, Tasmania 7001, Australia \\
$^7$ Fundaci\'on Galileo Galilei - INAF, Rambla Jos\'e Ana Fern\'andez P\'erez, 7, 38712 Bre\~na Baja, TF - Spain \\
$^8$ INAF-Osservatorio Astronomico di Palermo, Piazza del Parlamento 1, 90134 Palermo, Italy\\
$^9$ Lowell Observatory, 1400 W. Mars Hill Road, Flagstaff, AZ, 86001 USA \\ 
}
\date{}
\maketitle

\begin{abstract}

HD~179949 is an F8V star, orbited by a giant planet at $\sim 8~R_{\star}$ every $3.092514$~days. The system was reported to undergo episodes of stellar activity enhancement modulated by the orbital period, interpreted as caused by Star-Planet Interactions (SPIs). One possible cause of SPIs is the large-scale magnetic field of the host star in which the close-in giant planet orbits.

In this paper we present spectropolarimetric observations of HD~179949 during two observing campaigns (2009~September and 2007~June). We detect a weak large-scale magnetic field of a few Gauss at the surface of the star. The field configuration is mainly poloidal at both observing epochs. The star is found to rotate differentially, with a surface rotation shear of $\dom=0.216 \pm 0.061~\rpd$, corresponding to equatorial and polar rotation periods of $7.62 \pm 0.07$ and $10.3 \pm 0.8$~d respectively. The coronal field estimated by extrapolating the surface maps resembles a dipole tilted at $\sim70\degr$. We also find that the chromospheric activity of HD~179949 is mainly modulated by the rotation of the star, with two clear maxima per rotation period as expected from a highly tilted magnetosphere. In September~2009, we find that the activity of HD~179949 shows hints of low amplitude fluctuations with a period close to the beat period of the system.

\end{abstract}


\section{Introduction}

\label{sec:intro}

Hot Jupiters (HJs) are giant planets, orbiting close to their host stars (semi-major axis $< 0.1$ AU). They represent about 25\% of all discovered extrasolar planets. Interactions between the star and the planet can occur in such systems. Star-planet interactions (SPIs) can be of different types: tidal interactions due to the proximity and masses of the two bodies, and plasma interactions due to the magnetic field, e.g. reconnections between stellar and planetary fields \citep{rubenstein00,cuntz00,griessmeier07AA}. \cite{cuntz00} suggested that such interactions may enhance stellar activity, this enhancement being modulated by half of the orbital period in the case of tidal interactions, and by the orbital period in the case of magnetospheric interactions. Recently, \cite{fares10} suggested that the enhancement due to magnetospheric SPIs is more likely to be modulated on the beat period (synodic period between the stellar rotation and orbital periods) of the system. 

Observational studies of HJ hosting stars reported that not all observed systems show hints of interaction; moreover, for a single system, activity enhancement may be present at some epochs, yet absent at other epochs; finally, a phase lag between the subplanetary longitude and the peak of the activity enhancement is reported in some systems \citep{shk03,shk05,shk08}. Different theoretical studies have tried to explain these phase lags. \cite{mcivor06} considered the case where the stellar field is a tilted dipole, \cite{preusse06} adopted a model based on the propagation of Alfv\'en waves within the stellar wind flow relative to the planet and \cite{lanza08} considered a non-potential magnetic field configuration for the closed corona. \cite{cranmer07} modelled the \caii light curve of a HJ hosting star for different configurations of the stellar magnetic field (having real solar configurations throughout 11 year cycle), and showed that the  presence/absence of activity enhancement depends on the magnetic field geometry. A statistical survey of X-ray emission of stars with HJ suggests that they may be $\sim$~4 times more active than stars with distant planets \citep{kashyap08}, but this result was recently contradicted by \cite{poppenhaeger10} who found no significant correlations of X-ray luminosity with the planetary parameters in their sample of stars. Magnetohydrodynamic (MHD) simulations of \cite{cohen09} show that SPI may increase the X-ray luminosity.


HD~179949 is one of the most studied stars for SPI. It is an F8 star, orbited by a giant planet with minimum mass $ m \sin i = 0.916\pm0.076 M_{\rm \jupiter}$, semi-major axis $a = 0.0443\pm0.0026$~AU and orbital period $P_{orb}=3.092514\pm0.000032$~d \citep{butler06}. The star is reported to undergo epochs of activity enhancement modulated by the orbital period, with a phase lag of $\sim65\degr$ \citep{shk08}. We organised a multi-wavelength campaign to observe this star in 2009~September, almost simultaneously in spectroscopy (\'{e}chelle spectrograph~@~du Pont telescope~@~Las Campanas, PHOENIX~@~Gemini and SARG~@~TNG), spectropolarimetry (ESPaDOnS~@~CFHT) and X-rays (XMM-Newton). With such campaigns one can characterize the system, from the stellar surface (activity and magnetic field) to the stellar corona (X-rays). Simultaneous campaigns also provide an accurate description of the large-scale magnetic field allowing a quantitative modeling of the star-planet interaction.

In this first paper, we present the results of the spectropolarimetric campaign of 2009~September and of a previous campaign in 2007~June. We describe our data in section \ref{sec:obs}. The magnetic field reconstruction procedure, the magnetic properties and differential rotation of the star are presented in section \ref{sec:magnetic}, followed by coronal field extrapolation and activity analysis in section \ref{sec:extrapol} and \ref{sec:activity}. We summarize the results and discuss their implications in section \ref{sec:discussion}.


\section{Spectropolarimetric observations}
\label{sec:obs}
%
 
In order to map the surface magnetic field of HD~179949 and its evolution, we collected spectropolarimetric data during two observing campaigns in 2009~September and 2007~June. The data set collected in 2009~September is more extensive and of better quality than that collected in 2007~June - hence we analyze it first in the rest of the study. We used ESPaDOnS, a high-resolution spectropolarimeter installed at the 3.6-m Canada-France-Hawaii Telescope (CFHT) in Hawaii. ESPaDOnS provides spectra that span the whole optical domain (370 to 1000~nm) with a resolution of about $65000$ (when used in spectropolarimetric mode). 

The data were reduced using a fully automatic tool, called Libre-ESpRIT, installed at the CFHT for the use of the observers \citep{donati97}. From collected calibration exposures and stellar frames, Libre-ESpRIT automatically extracts wavelength calibrated intensity and polarisation spectra with associated error bars at each wavelength pixel. Each spectrum is extracted from four subexposures, taken in different configurations of the polarimeter retarders, in order to perform a full circular polarization analysis \citep{donati97}. Although it is possible to extract polarisation spectra from two subexposures, using four permits to eliminate (to first order) all systematic error or spurious polarisation (as well as producing a null-polarisation-test profile, to be discussed below). The spectra are normalised to a unit continuum, their wavelength scale refers to the Heliocentric rest frame. Telluric lines serve as a reference to correct the spectral shifts resulting from instrumental effects (e.g. mechanical flexures, temperature or pressure variations). The radial velocity (RV) precision of the spectra, using this calibration procedure, is about $30$~\ms. The RV measurements (listed in Table \ref{tab:log}) are in good agreement with the expectations when using the orbital solution of \cite{butler06}. Fig.~\ref{fig:RV} represents the RV variations. 

In 2009~September, we collected 19 spectra over 15 nights (two stellar rotations). Our data have good S/N ratio, ranging from 940 to 1910 per 2.6~\kms\ velocity bin around 700~nm; they sample well the rotation phases. In 2007~June, we collected 10 spectra, covering about $1.5$~rotational cycles. The S/N of the spectra around 700~nm ranges from 950 to 1280. The complete log of the observations is listed in Table \ref{tab:log}.

The rotational and orbital phases, denoted $\rm{E_{\rm Rot}}$ and $\rm{E_{\rm Orb}}$, were computed using the two ephemerides:
\begin{eqnarray}
T_0 &=& \mbox{HJD~}2,451001.51 + 3.092514~\rm{E_{\rm Orb}} \nonumber\\
T_0 &=& \mbox{HJD~}2,451001.51 + 7.6~\rm{E_{\rm Rot}}
\label{eq:eph}
\end{eqnarray}
The first ephemeris is that of \cite{butler06}, phase zero corresponding to the inferior conjunction (i.e. the planet being between the star and the observer). For the ephemeris giving the rotation phase, we use a rotation period of 7.6~d, identified as the equatorial rotation period (see section \ref{sec:magnetic}).  

\begin{table*}
\caption[]{Journal of September~2009 and June~2007 observations. Columns 1--8 sequentially list the UT date, the heliocentric Julian date (at mid-exposure), the complete exposure time, the peak signal to noise ratio (per 2.6~\kms\ velocity bin) of each observation (around 700~nm), the rotational and orbital cycles (using the ephemeris given by Eq.~\ref{eq:eph}), the radial velocity (RV) associated with each exposure, the rms noise level (relative to the unpolarized continuum level
$I_{\rm c}$ and per 1.8~\kms\ velocity bin) in the circular polarization profile produced by Least-Squares Deconvolution (LSD) and the false-alarm probability of the detection of the magnetic signature.}
\begin{tabular}{cccccccccc}
\hline
\hline
Date (UT) & HJD  & $t_{\rm exp}$  & \sn & Rot. Cycle & Orb. Cycle & \vrad & $\sigma_{\rm LSD}$ & fap\\
\hline
 2009   & (2455090+)  & (s) & & (539+) & (1325+)&(\kms) &(\ptt) & \\
\hline
25 Sep & 9.73903 & 4$\times$1260& 1750 & 0.2407 & 0.2095 &$-24.58$ & 0.21 & $<10^{-8}$\\
25 Sep & 9.80652 & 4$\times$1190& 1620 & 0.2495 & 0.2314 &$-24.57$ & 0.24 & $<10^{-8}$\\
27 Sep & 11.72735 & 4$\times$1260 & 1830 & 0.5023 & 0.8525 &$-24.39$ & 0.20 & $10^{-06}$\\
27 Sep & 11.79302 & 4$\times$1190 & 1780 & 0.5109 & 0.8737 &$-24.39$ & 0.21 & $<10^{-8}$ \\
28 Sep & 12.73190 & 4$\times$1260 & 1350 & 0.6345 & 1.1773 &$-24.61$ & 0.27 & $0.08$\\
28 Sep & 12.79096 & 4$\times$1190 & 1410 & 0.6422 & 1.1964 &$-24.60$ & 0.25 & $0.008$\\
29 Sep & 13.73057 & 4$\times$1260 & 1910 & 0.7659 & 1.5002 &$-24.45$ & 0.19 &  $0.0002 $\\
29 Sep & 13.78955 & 4$\times$1190 & 1870 & 0.7736 & 1.5193 &$-24.44$ & 0.20 & $0.00003$\\
30 Sep & 14.73560 & 4$\times$1260 & 1810 & 0.8981 & 1.8252 &$-24.39$ & 0.21 & $0.4$\\
30 Sep & 14.79547 & 4$\times$1190 & 1790 & 0.9060 & 1.8446 &$-24.39$ & 0.21 & $0.1$\\
01 Oct & 15.73477 & 4$\times$1260 & 1630 & 1.0296 & 2.1483 &$-24.62$ & 0.23 & $0.6$\\
01 Oct & 15.79425 &  4$\times$1190 & 1410 & 1.0374 & 2.1676 &$-24.62$ & 0.28  & $0.5$\\
02 Oct & 16.74999 &  4$\times$1260 & 940 & 1.1632 & 2.4766 &$-24.56$ & 0.42 & $0.0005$\\
03 Oct & 17.71701 &  2$\times$1260 & 980 & 1.2904 & 2.7893 &$-24.40$ & 0.32 & $<10^{-6}$\\
04 Oct & 18.80356 &  4$\times$1260 & 1650 & 1.4334 & 3.1407 &$-24.55$ & 0.23 & $<10^{-6}$\\
05 Oct & 19.79055 &  4$\times$1190 & 1540 & 1.5632 & 3.4598 &$-24.50$ & 0.25 & $0.01$\\
07 Oct & 21.75237 &  4$\times$1260 & 1330 & 1.8214 & 4.0942 &$-24.51$ & 0.27&$0.007$ \\
10 Oct & 24.73615 &  4$\times$1260 & 1650 & 2.2140 & 5.0590 &$-24.49$ & 0.23 & $<10^{-8}$\\
10 Oct & 24.79574 &  4$\times$1190 & 1480 & 2.2218 & 5.0783 &$-24.51$ & 0.26 & $<10^{-8}$\\
\hline
 2007   & (2454270+) &    & & (430+) & (1058+)& \\
\hline
23 June & 4.96537 &  4$\times$600 & 1280 & 0.7178 & 0.5095 &$-24.49$ & 0.30 & $0.03$\\
23 June & 5.08479 & 4$\times$600 & 1260 & 0.7335 & 0.5481 &$-24.44$ & 0.31 & $0.1$\\
26 June & 8.07869 & 4$\times$600 & 1060 & 1.1275 & 1.5162 &$-24.45$ & 0.38 &$0.4$\\
27 June & 8.87788 & 4$\times$600 & 1200 & 1.2326 & 1.7746 &$-24.36$ & 0.33 &$10^{-4}$\\
28 June & 9.87806 & 4$\times$540 & 950 & 1.3642 & 2.0980 &$-24.55$ & 0.43 &$0.9$ \\
01 July & 13.07675 & 4$\times$600 & 1000 & 1.7851 & 3.1324 &$-24.54$ & 0.42 & $0.2$\\
02 July & 14.09877 & 4$\times$600 & 990 & 1.9196 & 3.4629 &$-24.50$ & 0.43 & $0.003$ \\
03 July & 15.00800 & 4$\times$600 & 1050 & 2.0392 & 3.7569 &$-24.38$ & 0.38 & $0.2$\\
03 July & 15.07963 & 4$\times$600 & 900 & 2.0486 & 3.7800 &$-24.37$ & 0.47 & $0.3$\\
04 July & 15.97775 & 4$\times$700 & 870 & 2.1668 & 4.0705 &$-24.52$ & 0.47 & $0.5$\\
\hline
\hline
\end{tabular}
\label{tab:log}
\end{table*}

We apply Least-Squares Deconvolution (LSD) to our spectra in order to improve their S/N ratio \citep{donati97}. LSD consists of deconvolving each spectrum by a line mask, computed using a Kurucz model atmosphere with solar abundances, temperature of $6250$~K (following \citealt{santos04}) and logarithmic gravity of $4.0~\hbox{cm\,s$^{-2}$}$. The line mask includes the most moderate to strong lines present in the optical domain (those featuring central depths larger than 40\% of the local continuum, before any macroturbulent or rotational broadening, about 4,000 lines throughout the whole spectral range) but excludes the strongest, broadest features, such as Balmer lines, whose Zeeman signature is strongly smeared out compared to those of narrow lines. The contribution of each line to the average line profile is weighted by its depth, so that the strongest lines determine the shape of the LSD profile. The cutoff value of 40\% was selected at the start of the series of studies of planet-hosting stars \citep{moutou07,donati08,fares09,fares10} by optimising SNR in the LSD profile as a function of line-depth cutoff threshold, using the method of \cite{shorlin02}. In addition to the intensity and polarization profiles, LSD produces a null profile (labelled N) used as a polarization check and that should show no signal; this helps to confirm that the detected polarization is real and not due to spurious instrumental or reduction effects \citep{donati97}. None of the null profiles show a polarisation signature. The multiplex gain provided by LSD in V and N spectra is of the order of 25 with respect to a single line with average magnetic sensitivity, implying noise levels as low as 20 parts per million (ppm).

\begin{figure}
\includegraphics[height=.25\textheight]{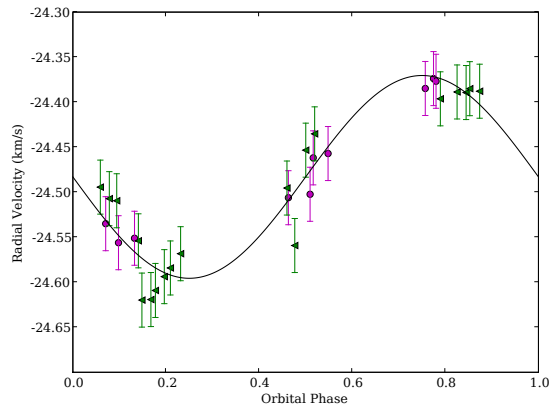}
\caption[]{Radial velocities of HD~179949 derived from September~2009 (green triangles) and June~2007 (magenta dots) spectra as a function of orbital phase, with their error bars. The radial velocity curve plotted here is that using the orbital period and amplitude from \cite{butler06}. }
\label{fig:RV}
\end{figure}


\section{Magnetic field and differential rotation}
\label{sec:magnetic}
%
\subsection{Magnetic modelling}

To reconstruct the magnetic map of the star and estimate the surface differential rotation, we use a tomographic imaging technique, called Zeeman-Doppler Imaging (ZDI). ZDI assumes that profile variations are only due to rotational modulation and differential rotation, and can invert (in this context) series of circular polarization Stokes V profiles into the parent magnetic topology and get an estimate of differential rotation at the surface of the star. The magnetic field is described by its radial poloidal, non-radial poloidal and toroidal components, all expressed in terms of spherical-harmonic expansions. The expressions of $B_r$, $B_\theta$ and $B_\phi$ can be found in \citealt{donati06}; $\alpha_{l,m}$, $\beta_{l,m}$~and $\gamma_{l,m}$~are the spherical harmonics coefficients describing, respectively, the radial poloidal, non-radial poloidal and toroidal components of the magnetic field. This description of the magnetic field has many advantages: both simple and complex magnetic topologies can be reconstructed \citep{donati01}; the energy of the axisymmetric and non-axisymmetric modes, as well as of the poloidal and toroidal components is calculated directly from the coefficients of the spherical harmonics. The mean magnetic energy is calculated by the mean over the stellar surface of $B_r^2+B_\theta^2+B_\phi^2$. Given the low value of the projected rotational velocity \vsini\ ($7.02\pm0.50$~\kms\ \citealt{valenti05}, $6.3\pm0.9$~\kms\ \citealt{groot96}), the resolution at the surface of the star is limited. We therefore truncate the spherical-harmonic expansions to modes with $\ell \leq 6$. The highest-order spherical harmonics resolvable on a stellar surface typically have orders $\ell$, given by $\ell = max \left( \frac{2 \pi \vsini}{FWHM}, 4 \right) $, where FWHM is the full width of half maximum of the intrinsic profile. The highest value of $\ell$~is  closely related to the number of surface resolution elements around the stellar equator, as described in detail by \citealt{morin10}, Sect 3.2. The FWHM of the intrinsic profile of HD 179949 is 8 \kms\ (see below), while \vsini = 7 \kms, implying that modes of order $\ell= 5$~or 6 are resolvable.\footnote[1]{We performed a map reconstruction using $\ell_{max}=4$, the results are discussed in appendix \ref{lmax4}.} Whereas the longitudinal resolution depends mainly on the \vsini, the latitudinal resolution depends also on the inclination of the star and the phase coverage of the observations.

To compute synthetic circular polarization profiles, ZDI decomposes the surface of the star into 5000 grid cells of similar projected areas (at maximum visibility) and calculates the contribution of each grid cell to the reconstructed profile, given the RV of the cell, the field strength and orientation, the location of the cell and its projected area. Summing the contribution of all grid cells yields the synthetic profile at a given rotation phase. ZDI proceeds by iteratively comparing the synthetic profiles to the observed ones, until they match within the error bars. Since the inversion problem is ill-posed, ZDI uses the principles of Maximum-Entropy image reconstruction \citep{skilling84} to retrieve the simplest image compatible with the data. The form we use for the regularisation function is $S=-\sum_{l,m} {l(\alpha_{l,m}^{2}+\beta_{l,m}^{2}+\gamma_{l,m}^{2})}$. More details about the currently used version of ZDI can be found in \cite{donati01,donati06}. Other details and assessment of the performance of an earlier version of ZDI code (which was not able to recover simple dipolar magnetic topology) can be found in \cite{brown91,donati97b}, as well as in appendix \ref{testZDI} for the currently used version of ZDI.

The models we use to describe the local unpolarized Stokes I and circular polarized Stokes V profiles at each grid cell are quite simple. Stokes I profiles are modelled by a Gaussian with FWHM of 8~\kms and central rest wavelength of 550~nm, this Gaussian FWHM fitting best the observations. The linear limb-darkening coefficient is set to $0.6$. Stokes V profiles are modelled assuming the weak field approximation, i.e. that the local Stokes V profile is proportional to the effective Land\'e factor (set to $1.2$), the line-of-sight projected component of the magnetic field and the derivative of the local Stokes I profile. This approximation is valid for HD~179949 (whose field strength is a few Gauss). The inclination angle of the star is approximately 60\degr\ given the \vsini\ of 7~\kms\ \citep{valenti05}, the stellar radius of $\rm 1.19\pm0.03~R_{\odot}$ \citep{fischer05} and the rotation period of $7.6$~d (see section \ref{sec:DR}). 

Differential rotation (DR) at the surface of the star can be estimated using ZDI. Magnetic regions at different latitudes have different angular velocities, their signatures in the spectra repeat with different recurrence rates. We assume that the rotation at the surface of the star follows $\Omega(\theta) = \omeq - \dom  \sin^2(\theta)$, where $\Omega(\theta)$ and \omeq\ are respectively the angular velocities at a latitude $\theta$~and at the equator, and \dom\ is the difference in the rotation rate between the pole and the equator. When the data cover more than a rotational cycle and are well sampled over the rotation cycle, one can in principle measure the recurrence rate of magnetic regions and thus deduce the differential rotation rate of the star. In practice, this is done by reconstructing a magnetic map for each pair of differential rotation parameters (\omeq, \dom) at a given information content (constant magnetic energy), and deriving the associated reduced chi-square \chisqr\ at which modelled spectra fit observations. The optimum DR parameters are the ones minimizing \chisqr. They are obtained by fitting the surface of the \chisqr~map with a paraboloid around the minimum value of \chisqr\ \citep{donati03}. 

\subsection{Results}

\subsubsection{Differential Rotation}
\label{sec:DR}
To determine the DR of HD~179949, we applied the method described above to 2009~September data (better sampled and covering a longer time span than the 2007~June data set). The \chisqr\ map we obtain forms a well-defined paraboloid shown in Figure~\ref{fig:DR}. The DR parameters we derive are $\omeq=0.824\pm0.007~\rpd$ and $\dom=0.216 \pm 0.061~\rpd$, implying an equatorial rotation period of $7.62 \pm 0.07$~d and a polar rotation period of $10.3 \pm 0.8$~d. The equatorial rotation period we find is compatible with the one found by \citealt{shk03} ($\le 9$~ days) but slightly higher than the one weakly detected by \citealt{wolf04} (7.06 days). \cite{cameron07} describes \dom\ as a power-law function of the stellar effective temperature; the derived \dom\ for HD~179949 complies with this power law.


\begin{figure}
\includegraphics[angle=-90,scale=0.35]{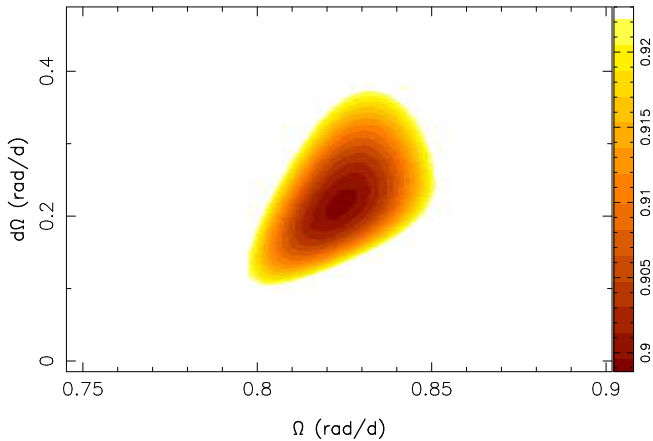}
\caption[]{Variations of \chisqr\ as a function of \omeq\ and \dom\ , derived from the modelling of the Stokes V data set for 2009 September. The outer colour contour corresponds to a $2.7\%$~increase in the \chisqr, and traces a $3~\sigma$~interval ($99.7 \%$~confidence level) for both parameters taken as a pair.}
\label{fig:DR}
\end{figure}

For 2007~June data set, we were not able to measure the DR parameters. The data are more noisy than those of 2009~September and the sampling of the rotational cycle is poorer. If we fit the data down to a \chisqr~value of $0.75$ (which we consider normally as overfitting the data), we get the DR parameters similar to those of 2009~September.

\subsubsection{Magnetic configurations}
\label{sec:maps}
For 2009~September, the reconstructed Stokes V profiles fit well the data down to a \chisqr\ value of 0.9 (see Fig. \ref{fig:StokesV}). The surface magnetic field (Fig. \ref{fig:maps}), producing these circular polarization profiles, is mainly poloidal (the poloidal energy contributes 90\% of the total energy). Low order spherical harmonics contribute the most of the poloidal energy, the dipolar and quadrupolar contribution being about 40\%\ each. 40\%\ of the poloidal field is in axisymmetric modes (note that non-axisymmetric modes correspond to those with $m \geq l/2$ and axisymmetric ones to those with $m < l/2$). The toroidal field constitutes a small fraction of the total field (about 10\%).


\begin{figure*}
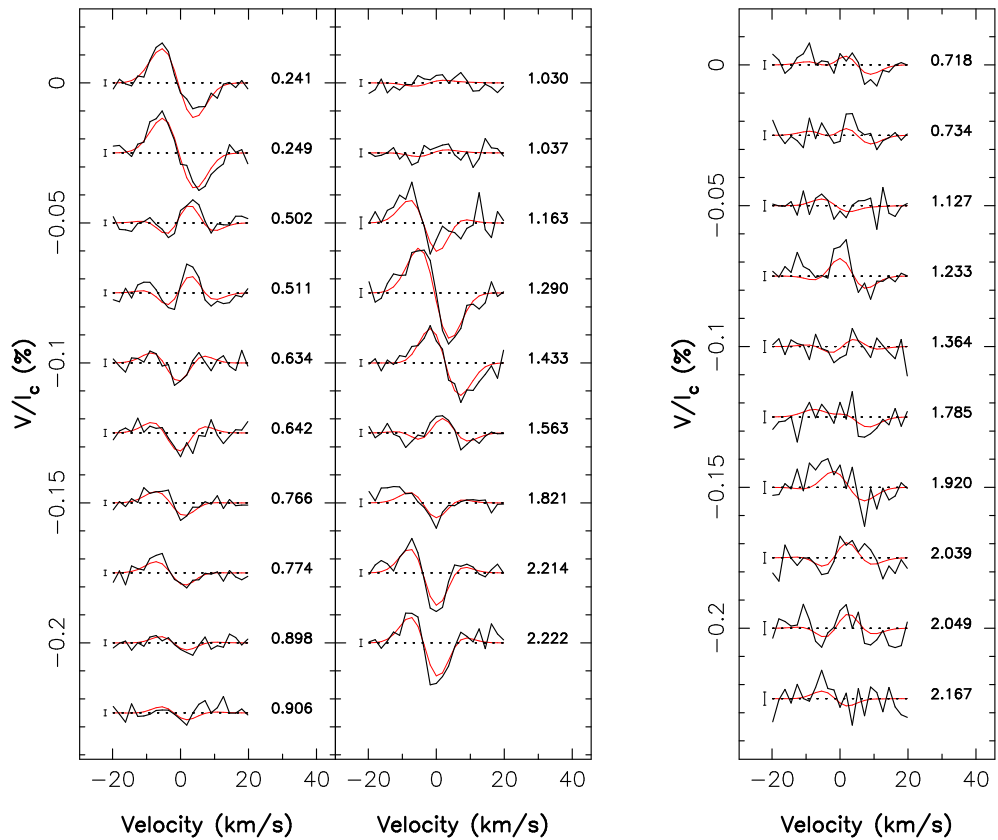

\center{\hbox{\includegraphics[angle=-90,scale=0.6]{ProfilV_sep09_paper.ps}\hspace{10mm}\includegraphics[angle=-90,scale=0.6]{ProfilV_jun07_paper.ps}}}
\caption[]{Circular polarization profiles of HD~179949 for 2009~September (left panel) and 2007~June (right panel). The observed and synthetic profiles are shown in black and red respectively. On the left of each profile we show a $\pm1~\sigma$ error bar, while on the right the rotational cycles are indicated.}
\label{fig:StokesV}
\end{figure*}


For June~2007, assuming that the DR did not change over the two epochs, we fit the observed profiles to a \chisqr\ value of 0.9. The field is weaker than September~2009, but is still mainly poloidal (80\%\ of the total energy). The poloidal component is mostly axisymmetric (60\%). A summary of the field properties is presented in Table \ref{tab:field}. Due to the low S/N values we have for June~2007 data, the reconstructed magnetic energy could be less than the true energy. A Discussion of the effect of S/N and phase sampling on the reconstruction can be found in \cite{donati97b} and in appendix \ref{testZDI}.

\begin{table*}
\caption[]{The field properties for the two epochs of observations. The magnetic field strength averaged over the stellar surface is indicated in the second column (the error bar is of $\pm 0.3$~Gauss for both seasons). The percentage of the poloidal energy relative to the total one and the percentage of the axisymmetric modes (modes with $m<l/2$) energy in the poloidal component relative to the poloidal energy are reported in column 3 and 4 respectively. The energy in the modes having $m=0$ is reported in column 5. }
\begin{tabular}{ccccc}
\hline
Epoch& B& $\rm E_{poloidal}$ & $\rm E_{axisymmetric, m<l/2}$ & $\rm E_{m=0}$\\
& (G) & \% of $\rm  E_{total}$& \% of poloidal &  \% of poloidal\\
\hline
September~2009& 3.7 & 90  & 36 & 35 \\
\hline
June~2007 & 2.6 & 80  & 55 & 54 \\
\hline
\end{tabular}
\label{tab:field}
\end{table*}

\begin{figure*}
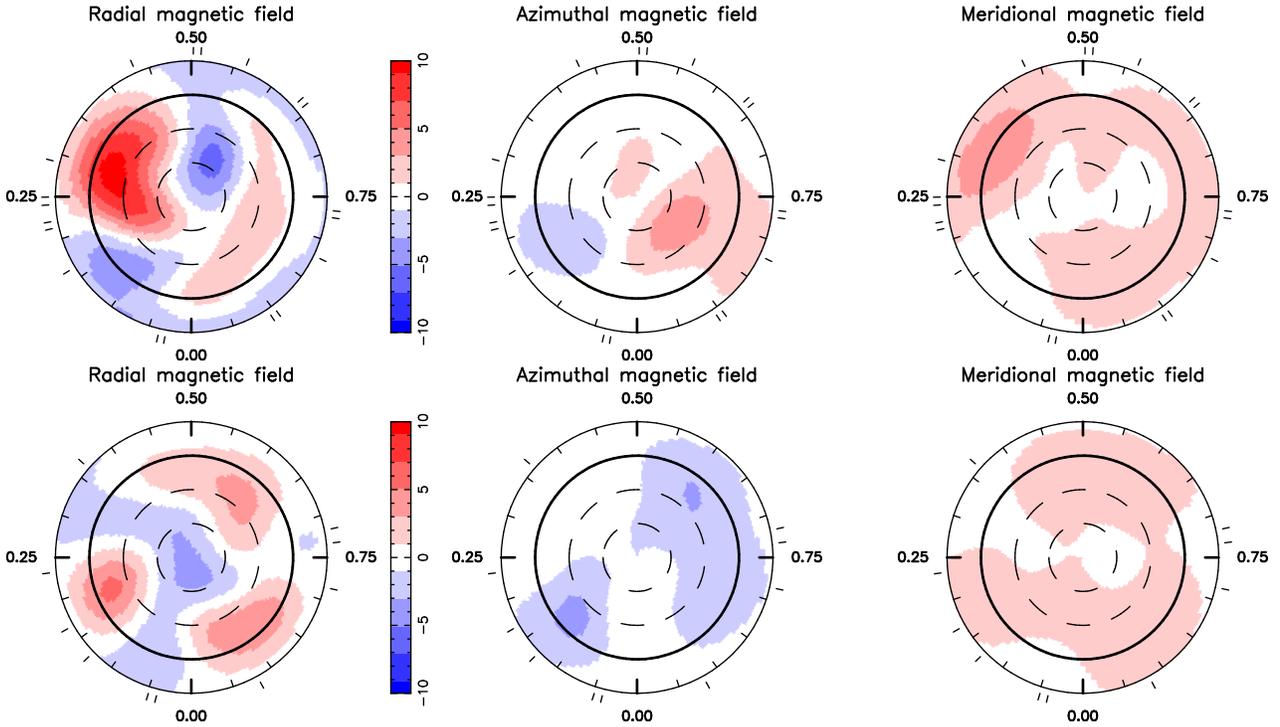

\center{\includegraphics[scale=0.7]{Map_sep09_paper_largeur.ps}
	\includegraphics[scale=0.7]{Map_jun07_paper_largeur.ps}}
\caption[]{Magnetic maps of HD~179949 for 2009~September (upper panel) and June~2007 (lower panel). The three components of the field in spherical coordinates system in a flattened polar view of the star are presented, down to latitude -30\degr\ . The bold circle represents the equator. The small radial ticks around the star represent the rotational phases of our observations. The radial, azimuthal and meridional fields have the same color scale.}
\label{fig:maps}
\end{figure*}
Comparing the magnetic field at both epochs, we notice that it is slightly stronger ($\sim 1$~G) in 2009~September than in 2007~June, but this can be due to the better quality of the data set. Cool stars having a Rossby number and a mass greater than $1$~and $0.5~\mSun$ respectively seem to have the same trend of magnetic configurations, i.e. mainly axisymmetric poloidal field \citep[Figure 3]{donati09}. In this context, the large-scale field configuration of HD~179949 is typical for stars with similar masses and rotation rates.

\section{Large-scale coronal magnetic field}
\label{sec:extrapol}

Using the surface magnetic maps, we can extrapolate the magnetic field to the corona. In this end, we use the Potential Field Source Surface (PFSS) code. This code was first written to model the solar coronal field \citep{vanballegooijen98}, and was later developed to be used for stars other than the Sun \citep{jardine02}. 

We briefly summarise here the PFSS basic principles and underlying assumptions:
\begin{enumerate}
\item The magnetic field {\bf B} is written as the gradient of a scalar potential $\psi$ as ${\bf B} = - \nabla \psi$, i.e. the field is potential (${\bf \nabla \times B = 0}$)
\item The three components of the coronal field $B_{r}$, $B_{\theta}$ and $B_{\phi}$ are described by a spherical harmonics expansion
\item There is a source surface (of radius $R_{s}$) beyond which the field is purely radial $B_{\theta}(R_{s})= B_{\phi}(R_{s})=0$ 
\item The boundary conditions for the radial field at the stellar surface are given by the ZDI magnetic maps (when having these maps, otherwise the magnetic field is modelled by a theoretical model).
\end{enumerate}

Fig. \ref{fig:extrapol} shows as an illustrative example the coronal field lines for the image corresponding to 2009~September, for a source surface located at $3.4~R_{\star}$ (a plausible value given that the mean value of source surface radius of the Sun is $\sim 2.5~R_{\odot}$). Closed loops, connecting regions of different magnetic polarities, reach different heights in the corona ($< R_{S}$). Open field lines have a configuration close to those of a dipole tilted by $\sim70\degr$ relative to the rotational axis. The planet, at $8.1~R_{\star}$, is beyond the source surface. Its magnetic field can reconnect with open field lines, along which the stellar wind can be launched. Given the field geometry and the inclination of the star, the open field lines are almost oriented along the line of sight for some rotational phases (0.2 to 0.35). The large-scale coronal magnetic field in 2007~June is also shown in Figure \ref{fig:extrapol}.

\begin{figure*}
\center{\includegraphics[height=.25\textheight]{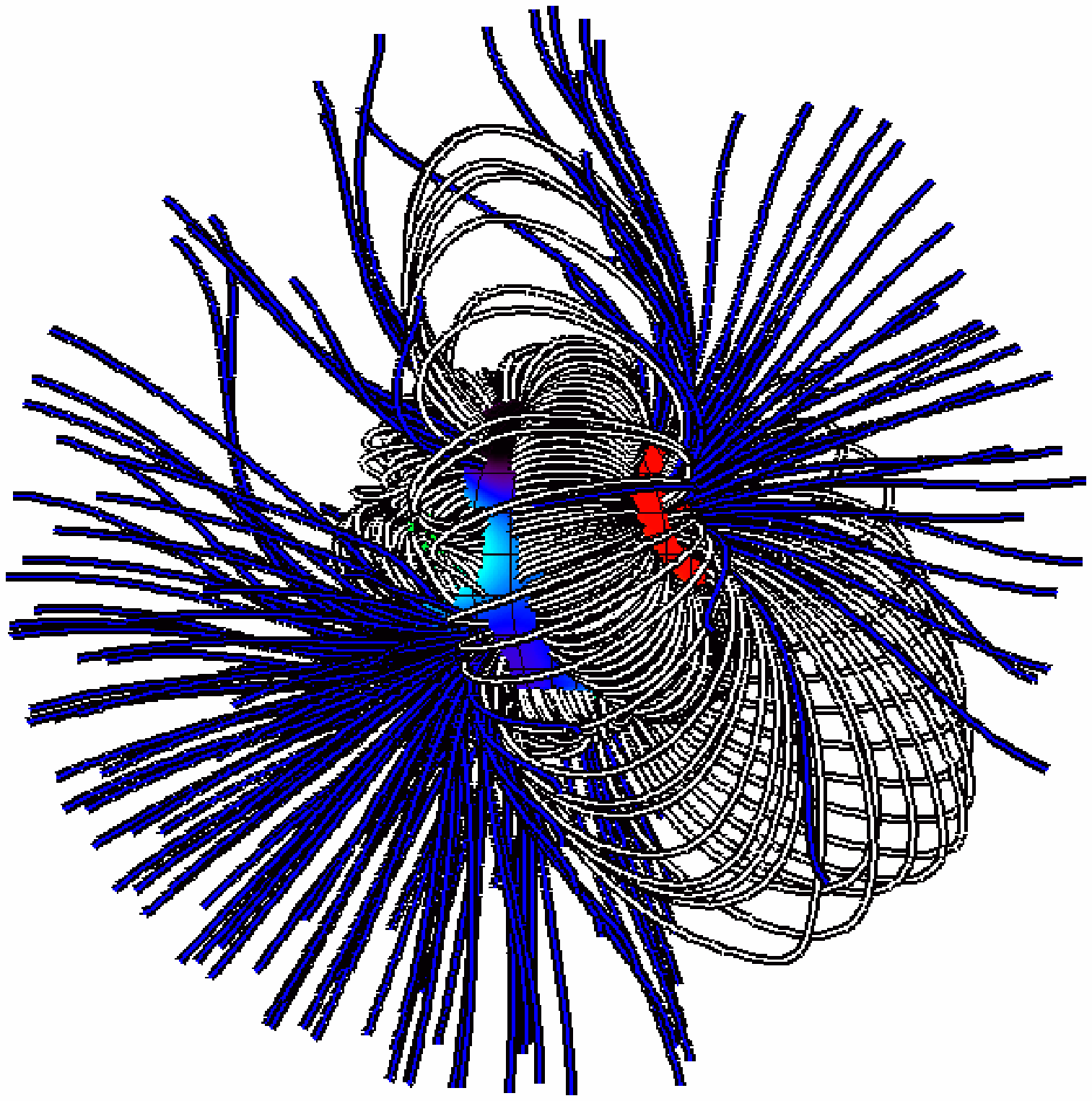}
        \includegraphics[height=.25\textheight]{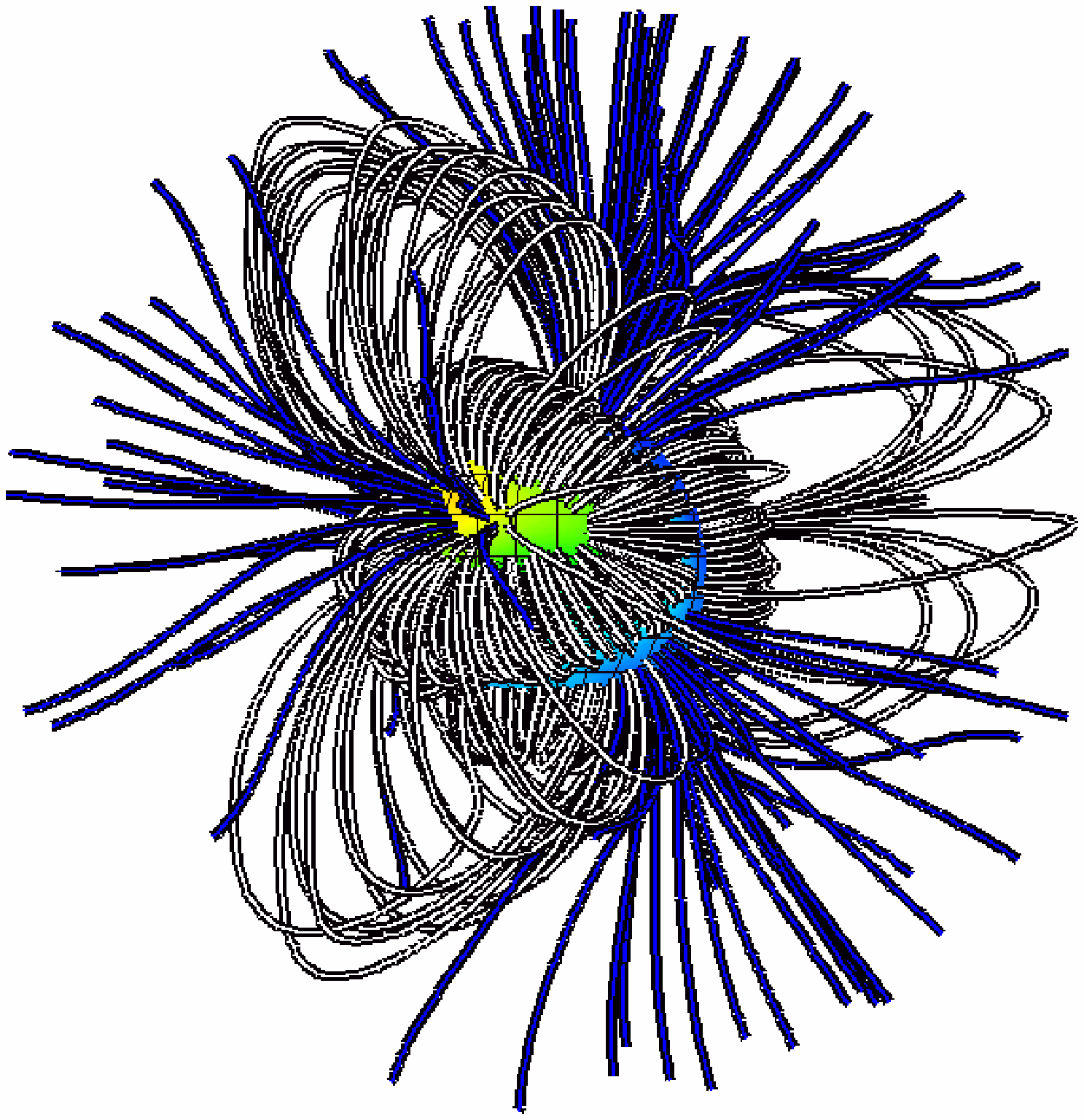}}
\caption[]{The extrapolated magnetic field of HD~179949 for 2009~September (left panel) and 2007~June (right panel). White lines corresponds to the closed magnetic lines and blue ones to the open field lines (reaching the source surface). The star is shown edged-on.}
\label{fig:extrapol}
\end{figure*}

The extrapolation of the coronal field is essential to understand the environment in which the HJ evolves. It shows for this star that the large scale magnetic field resembles a tilted dipole.


\section{stellar activity}
\label{sec:activity}

\begin{figure*}
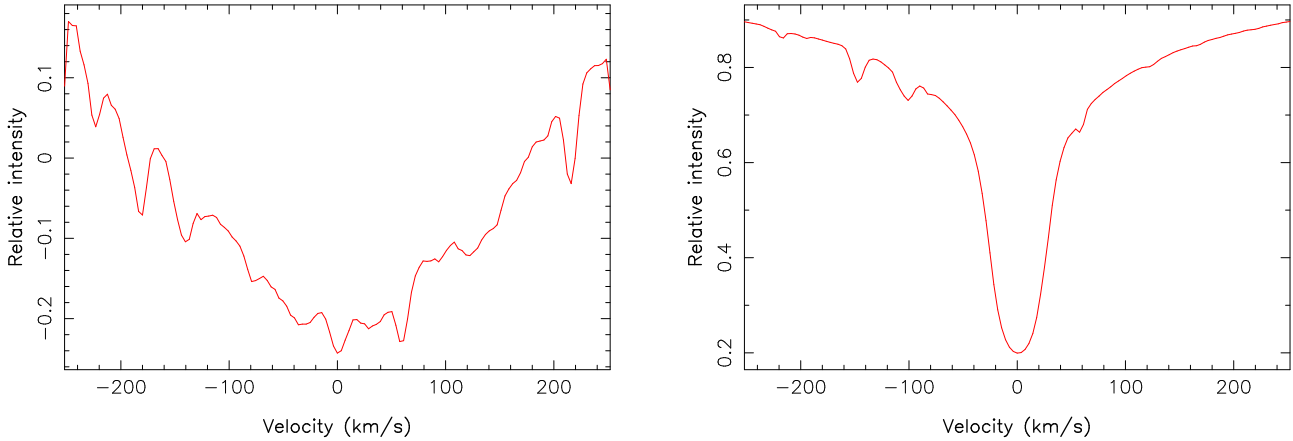

\center{\hbox{\includegraphics[angle=-90,scale=0.35]{mean_cah_sep09_referee.ps}\hspace{10mm}\includegraphics[angle=-90,scale=0.35]{mean_hal_sep09_referee.ps}}}
\caption[]{Mean \caii (left panel) and  H$\alpha$ (right panel) profiles for September~2009.}
\label{fig:mean_proxy}
\end{figure*}

\begin{figure*}
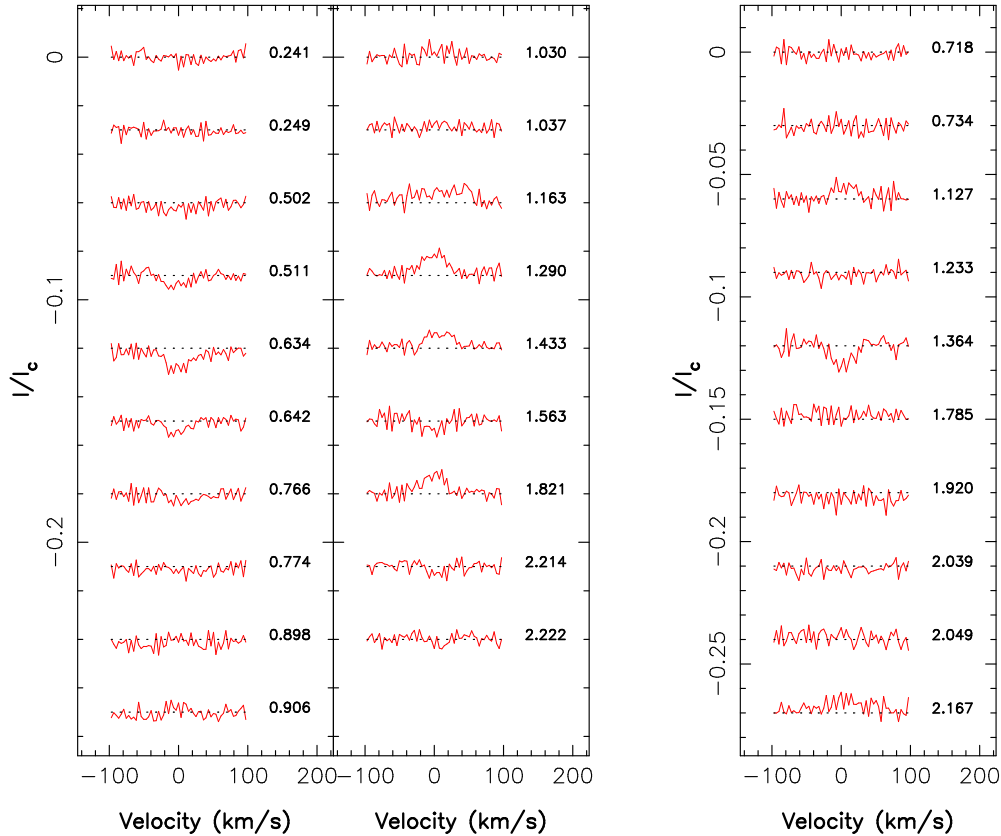

\center{\hbox{\includegraphics[angle=-90,scale=0.6]{Cah_sep09_19spectra.ps}\hspace{10mm}\includegraphics[angle=-90,scale=0.6]{Cah_jun07.ps}}}
\caption[]{The \caii residual emission in September~2009 (left panel) and June~2007 (right panel). They show variable features at the center of the line. The rotational phases are indicated on the right of each profile.}
\label{fig:activityresiduals}
\end{figure*}

To study the temporal evolution of chromospheric activity, we analyzed two chromospheric activity proxies : \caii and \halpha lines.

For each observing season (2009~September and 2007~June), we calculate a mean profile per proxy. Figure \ref{fig:mean_proxy} shows these mean profiles for 2009~September. The variability in the proxy is calculated by subtracting the mean profile to the observed ones. The core of the residual profiles vary with time (see Fig. \ref{fig:activityresiduals}). To estimate the residual emission, we fit the line core with a Gaussian (fixing its FWHM and its center at the line center); the equivalent width of the Gaussian gives the residual emission; the residual emission values are mentioned in Table \ref{tab:valeur}. The \caii and \halpha residual emission are clearly correlated (Fig. \ref{fig:correlation}, left panel).  

\begin{figure*}
\center{\hbox{\includegraphics[height=.25\textheight]{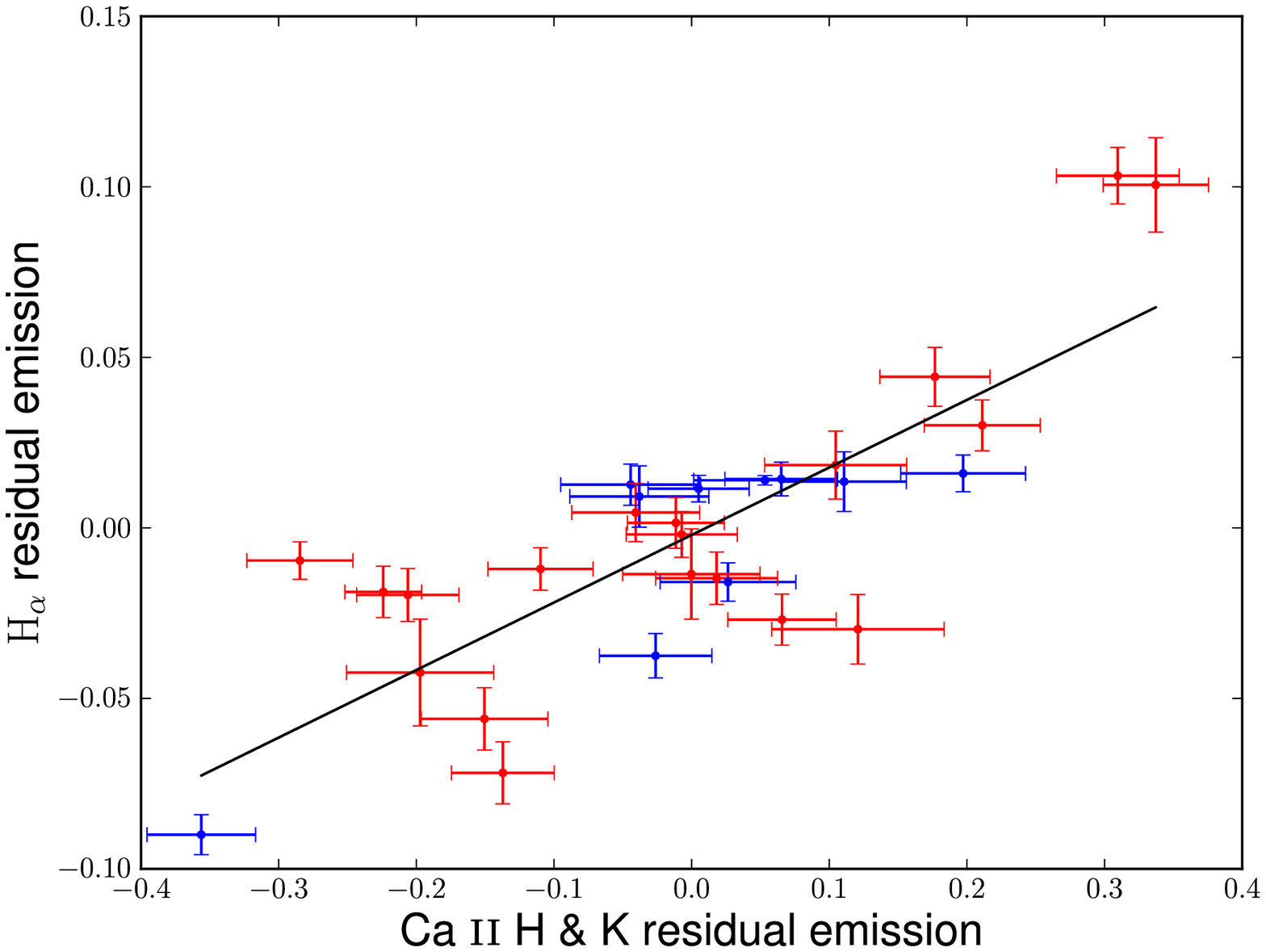}\includegraphics[height=.25\textheight]{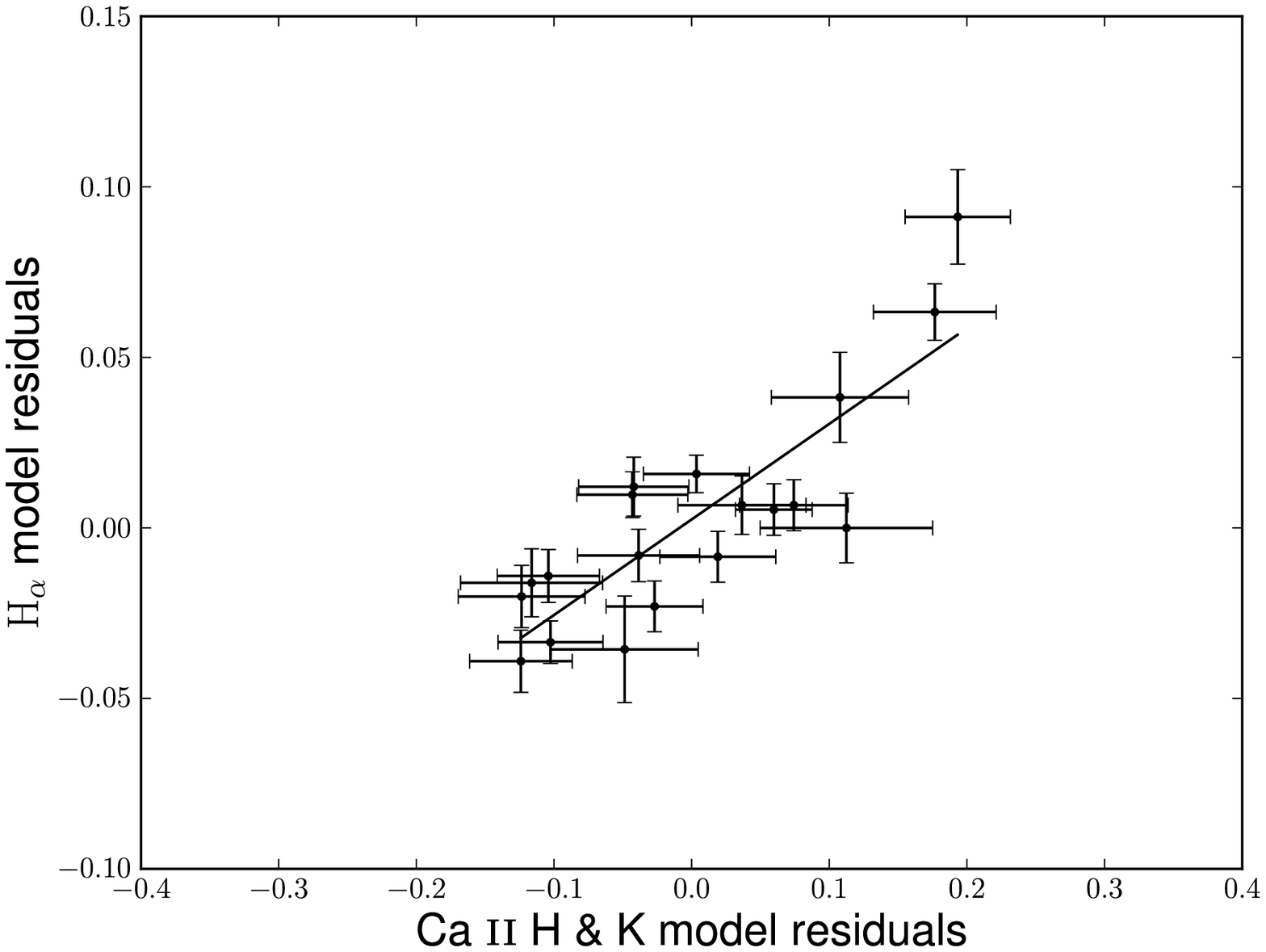}}}
\caption[]{Left panel: \halpha vs \caii residual emission for both epochs of observation (2009~September in red and 2007~June in blue). The two quantities are correlated, the Pearson correlation coefficient is $0.8$. The line is the best-fit linear function to the data, its slope is $0.20$. Right panel: \halpha and \caii residuals for 2009~September after subtracting the best-fit model to the data at $\rm P_{rot}$. The slope of the best-fit linear function is $0.28$.}
\label{fig:correlation}
\end{figure*}

The star exhibits variability on both short and long time scales. A modulation of the stellar activity by the orbital period was proposed as a proof of SPI magnetic interactions \citep{cuntz00}. In addition to a modulation with the orbital period, a modulation with the synodic period of the planet with respect to the stellar rotation is in principle possible. The synodic (or beat) period is the period for which the planet faces the same configuration of the magnetic field, and thus if this configuration is favourable for a magnetic interaction (e.g. reconnection), the interaction will happen again after a synodic period. Our aim is thus to search for a periodic modulation of the residual activity. We find that the activity is modulated, to first order, by the stellar rotation, and not by the beat period nor by the orbital period, see figure \ref{fig:calciumphase}. In order to search for low amplitude periodic fluctuation of the activity, two fitting approaches were used (a linear and a non-linear one). They will be detailed below.

\begin{figure*}
\center{\hbox{\includegraphics[height=.25\textheight]{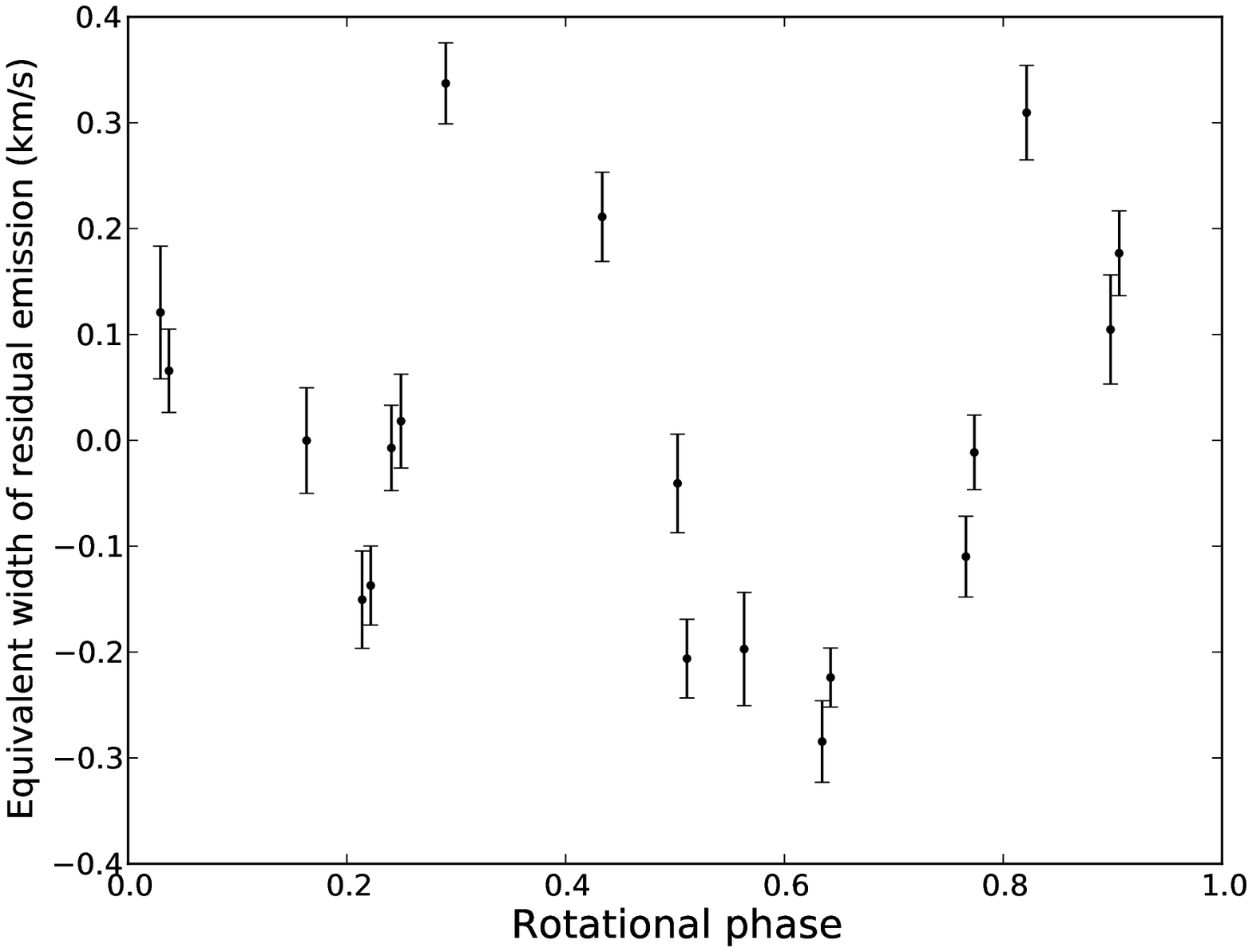}\includegraphics[height=.25\textheight]{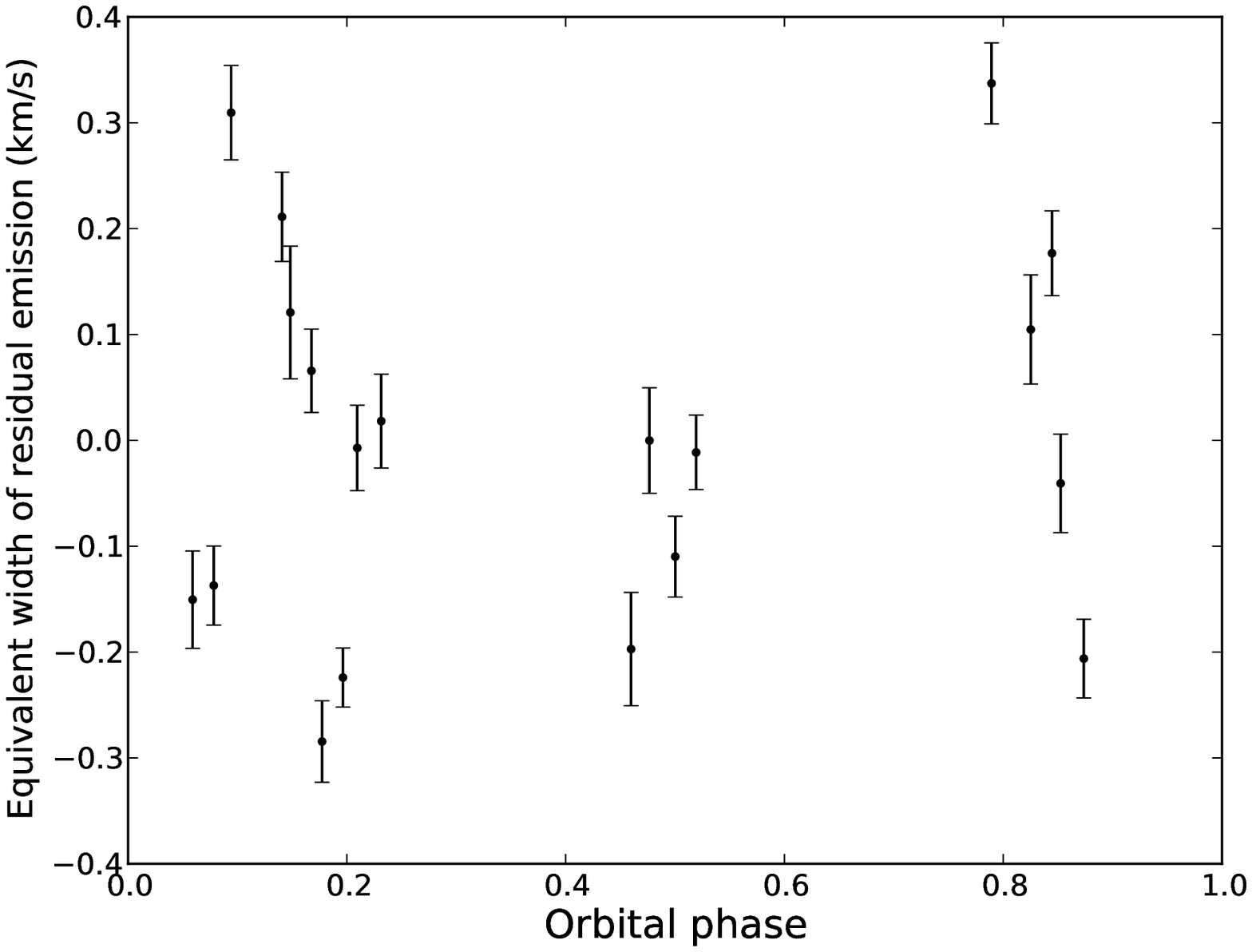}}}
\caption[]{The \caii residual emission in September~2009 as a function of the rotational phase (left panel) and orbital phase (right panel). }
\label{fig:calciumphase}
\end{figure*}

\subsection{Linear fit to the data}
The open field lines resemble those of a strongly tilted dipole (see section \ref{sec:extrapol}), causing the activity to be mainly modulated by half the rotation period. We model the variability by a sum of a linear function (for the long term variability) and two sine waves (one of period $P$ and one of period $P/2$). This model includes the main frequency of the data and its first harmonic (adding a third component at frequency $P/3$~as in \citealt{boisse11} does not modify the result significantly). Since the star rotates differentially, the rotational period ranges from $7.62 \pm 0.07$~d to $10.3 \pm 0.8$~d. For our fit, we fix the period and fit the other parameters of the model using a rejection criterion to omit the points distant by more than $10~\sigma$ to the model. We explore all the domain of rotational periods, repeating this procedure for every rotational period; the best-fit rotational period is the one minimizing the \chisqr\ of the fit (the number of rejected points never exceeds 2 per fit).

Since we are looking for a possible modulation by the beat period and since activity is modulated mainly by rotation, we first subtract the best fit model to the data and look for periodic fluctuations in the residuals (to be called model residuals not to be confused with the residual activity). We fit these model residuals by a sine wave, and calculate the best-fit period of these model residuals.

\begin{figure*}
\center{\hbox{\includegraphics[height=.25\textheight]{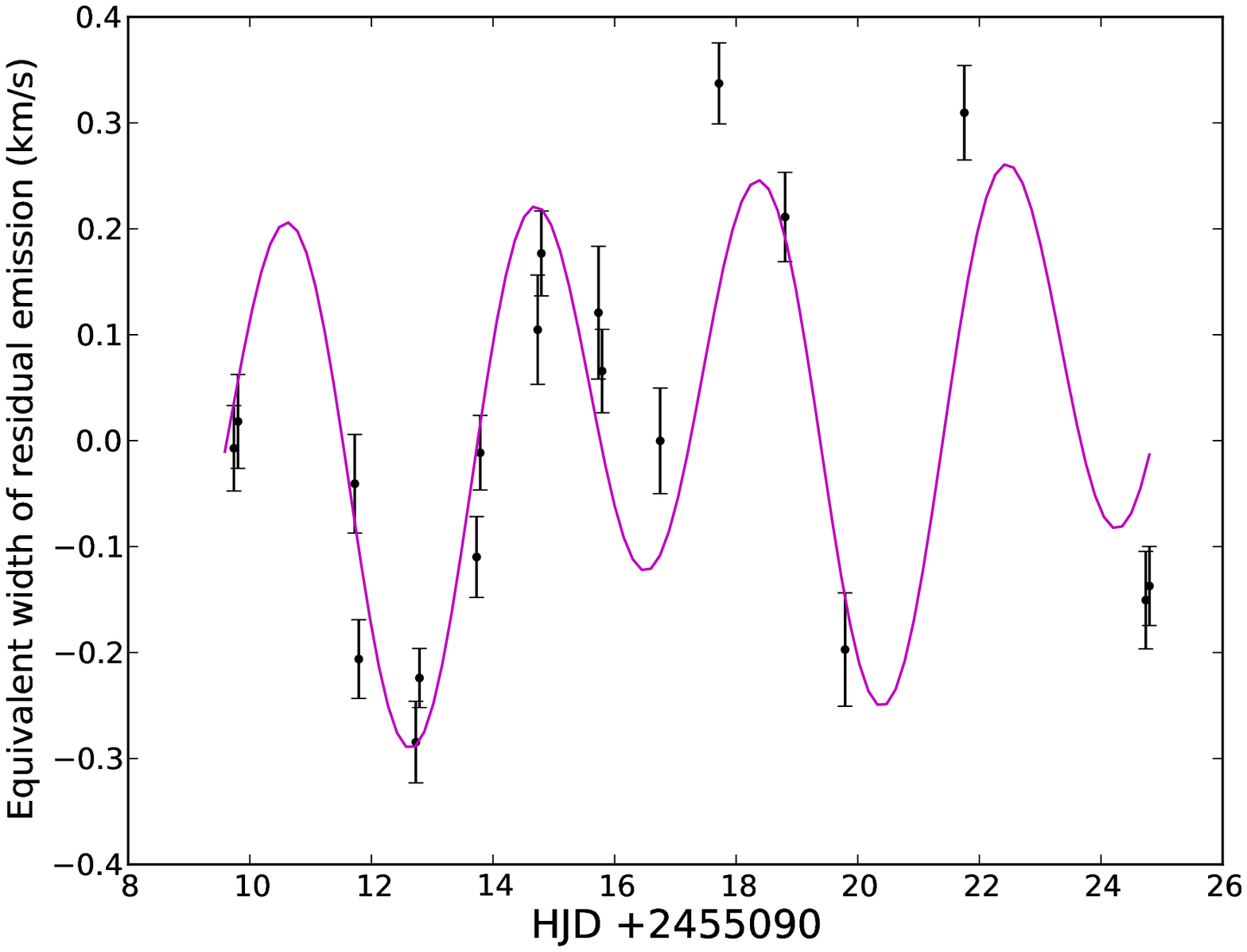}\includegraphics[height=.25\textheight]{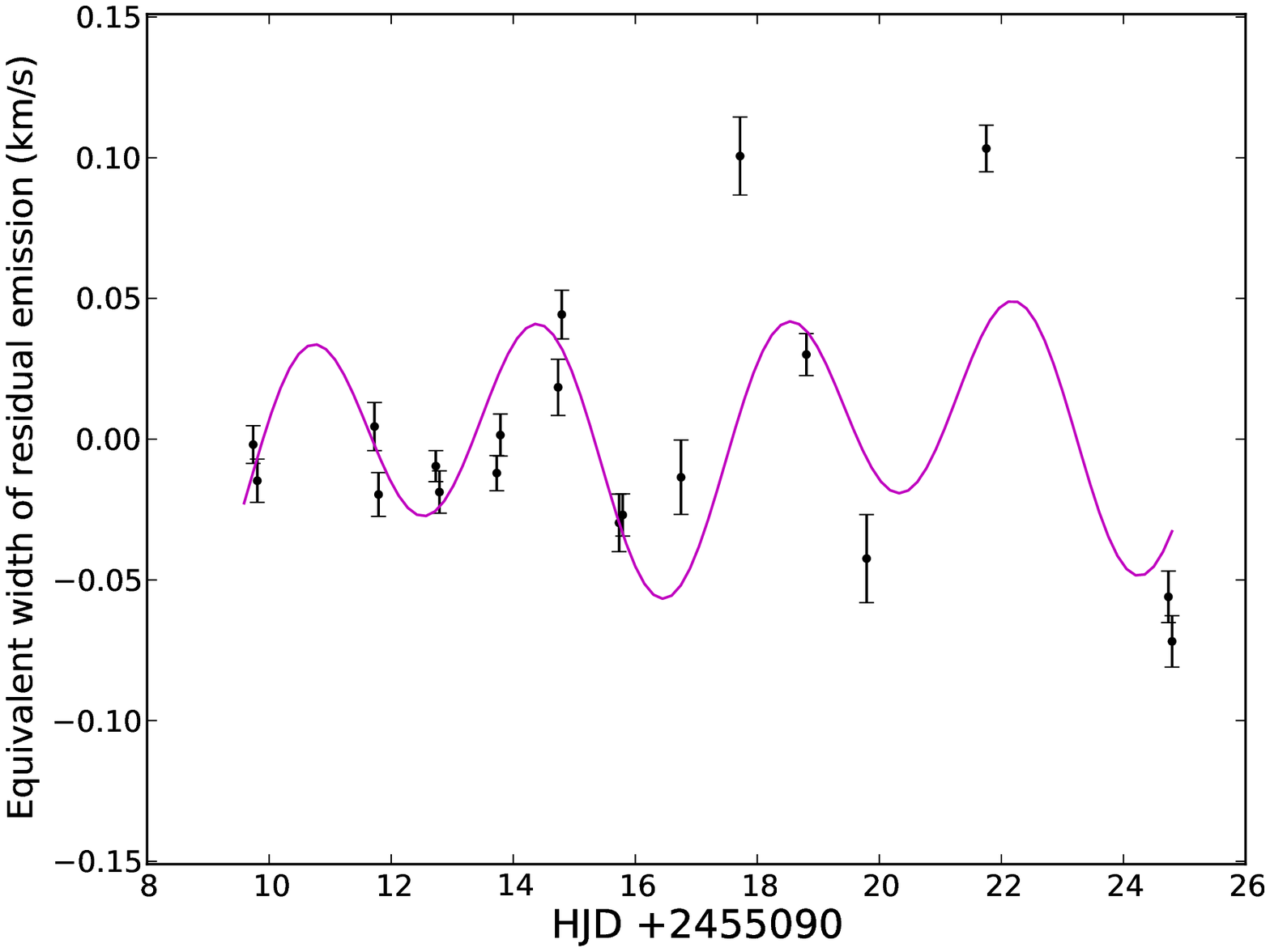}}}
\caption[]{\caii (left) and \halpha (right) residual emissions fitted by a linear function and a sum of two sine waves having as periods P and P/2 for September~2009. The best fit period to the data is 7.75~d and 7.8~d respectively.}
\label{fig:fitCa}
\end{figure*}

\begin{table}
\begin{center}
\begin{tabular}{ccc}
 HJD      & \caii  & \halpha \\
& residual emission &residual emission \\
\hline
$2455090+$ & (\kms)& (\kms)\\
\hline
9.73903 & $ -0.007 \pm 0.040 $&$ -0.002 \pm 0.007 $ \\
9.80652 & $ 0.018 \pm 0.044 $&$ -0.015 \pm 0.008 $ \\
11.72735 & $ -0.041 \pm 0.047 $&$ 0.004 \pm 0.009 $ \\
11.79302 & $ -0.206 \pm 0.037 $&$ -0.020 \pm 0.008 $ \\
12.73190 & $ -0.285 \pm 0.039 $&$ -0.010 \pm 0.005 $ \\
12.79096 & $ -0.224 \pm 0.028 $&$ -0.019 \pm 0.008 $ \\
13.73057 & $ -0.110 \pm 0.038 $&$ -0.012 \pm 0.006 $ \\
13.78955 & $ -0.011 \pm 0.035 $&$ 0.001 \pm 0.007 $ \\
14.73560 & $ 0.105 \pm 0.052 $&$ 0.018 \pm 0.010 $ \\
14.79547 & $ 0.177 \pm 0.040 $&$ 0.044 \pm 0.009 $ \\
15.73477 & $ 0.121 \pm 0.063 $&$ -0.030 \pm 0.010 $ \\
15.79425 & $ 0.066 \pm 0.039 $&$ -0.027 \pm 0.007 $ \\
16.74999 & $ 0.000 \pm 0.050 $&$ -0.014 \pm 0.013 $ \\
17.71701 & $ 0.337 \pm 0.038 $&$ 0.101 \pm 0.014 $ \\
18.80356 & $ 0.211 \pm 0.042 $&$ 0.030 \pm 0.007 $ \\
19.79055 & $ -0.197 \pm 0.053 $&$ -0.042 \pm 0.016 $ \\
21.75237 & $ 0.310 \pm 0.045 $&$ 0.103 \pm 0.008 $\\
24.73615 & $ -0.150 \pm 0.046 $&$ -0.056 \pm 0.009 $ \\
24.79574 & $ -0.137 \pm 0.037 $&$ -0.072 \pm 0.009 $ \\
\hline
$2454270+$ &&\\
\hline
4.96537 & $0.065 \pm 0.041$ & $0.014 \pm 0.005$\\
5.08479 &$0.053 \pm  0.052$  & $0.014 \pm 0.001$ \\
8.07869 &$0.197 \pm 0.045$ & $0.016 \pm 0.005$ \\
8.87788 &$-0.026 \pm 0.041$ & $-0.038 \pm 0.007$ \\
9.87806 & $-0.356 \pm 0.039$ & $-0.090 \pm 0.006$ \\
13.07675 & $-0.038 \pm 0.051$ & $0.009 \pm 0.009$ \\
14.09877 &$-0.044 \pm 0.051$ & $0.013 \pm 0.006$ \\
15.00800 & $0.005 \pm  0.037$ & $ 0.011 \pm 0.004 $ \\
15.07963 & $0.026 \pm 0.049$ & $-0.016 \pm 0.006 $\\
15.97775 &$0.111 \pm 0.045$ & $0.014 \pm 0.009 $\\
\hline
\end{tabular}
\caption[]{\caii and \halpha residual emission values for September~2009 (upper part of the table) and June~2007 (lower part of the table).}
\label{tab:valeur}
\end{center}
\end{table}

For 2009~September, the \caii and \halpha residual emissions are modulated by $7.75^{+0.55}_{-0.55}$~d and $7.8^{+1.15}_{-1}$~d respectively, reflecting the fact that both quantities are strongly correlated (see figure \ref{fig:fitCa} and left panel of figure \ref{fig:correlation}, and table \ref{tab:valeur} for the residual emission values). The slope of the best-fit linear function of \halpha vs \caii residual emission is $0.20$. To calculate the false-alarm probability (fap) of these signals, we produce 10000 simulated data sets by night-shuffling, and fit each data set using the same procedure as for our observed data set. The number of data sets for which the \chisqr\ of the best-fit period is inferior to the \chisqr\ of our original signal are divided by 10000 to give the fap (Note that we also used the gaussian random noise method to simulate the data sets; the calculated faps are similar to the ones calculated using night-shuffling). The fap of the \caii periodicity is about 6\%, that of \halpha is higher ($\sim 30\%$). 

We then subtracted the best-fit model for each proxy, and calculated the model residuals. The \caii model residuals are modulated by $5.3^{+0.6}_{-0.65}$~d, the \halpha ones by $5.5^{+0.7}_{-0.8}$~d, with a fap of 20 and 7\%\ respectively. These model residuals are correlated, as shown in Fig. \ref{fig:correlation} (right panel). The slope of the \halpha model residuals vs \caii model residuals is $0.28$, higher than that of the residual emission ($0.20$). This correlation may be interpreted as an activity enhancement, seen in both proxies, due to the same physical phenomenon. The beat period of the system ranges from 4.27 to 5.23~days, roughly compatible with the period of the observed fluctuations (within the error bars). The activity enhancement in the model residuals does not correspond to a single orbital phase. This enhancement happens at different orbital cycles (0.86, 2.47 and 4.1), contrary to what was already reported for this system.

For 2007~June, the \caii residual emission is modulated by $8.6^{+0.65}_{-3.5}$~d (fap of $0.06\%$). We do not find a periodic modulation in the model residuals of the \caii emission after subtracting the best-fit model. No hints of SPI are found for this epoch.

\subsection{Non-linear fit to the data}

We then used a second period search method, consisting of a non-linear fit of the data, using our original model : a linear function and three sine waves having as periods $P_{1}$, $P_{1}/2$~and $P_{2}$. We applied this method for the 2009~September data set. The periods we find for the \caii residual activity are $P_{1}=8.4$~days and $P_{2}=4.87$~days. These periods are compatible with the ones found using the first method within the error-bars. For the \halpha residual activity, the periods we found are $6.0$~and $1.27$~days.
 
Discrepancy between the results found using the two fitting methods may be due to the limited number of unevenly spaced data points, that was also suggested by the high fap probability we got using the first method. The source function of the \halpha line is dominated by the effects of photoionization and recombination in solar-like stars and F stars, so it is not mainly coupled to the local electron temperature in the chromosphere but principally to the UV radiation field of the star whose main component comes from the photosphere. The \caii lines are a more suitable proxy for the non-radiative heating produced by magnetic fields in the chromosphere of an F-type star. A definite detection/no-detection of low amplitude periodic fluctuation needs a larger data set.

\section{Summary}
\label{sec:discussion}

We present in this paper the first study of the large-scale magnetic field of the Hot Jupiter hosting star HD~179949.  We reconstructed the large-scale surface magnetic field at two observing epochs in 2009~September and 2007~June using ZDI. The field is mainly poloidal (the poloidal component contributing $80-90\%$ of the total energy), with a strength of a few Gauss (up to 10 Gauss in some local regions). The field configurations, properties, and polarity show no main changes when comparing both observing seasons. We detected differential rotation at the surface of the star, the latitudinal shear between the pole and the equator being of $\dom = 0.216 \pm 0.061~\rpd $ and the angular velocity at the equator being of $\omeq=0.824\pm0.007~\rpd$. The equatorial and polar rotation periods are of $7.62 \pm 0.07$~d and $10.3 \pm 0.8$~d respectively.
The overall field polarity is similar during both seasons, though we can not exclude an evolution on short timescale \citep{fares09}. We will continue monitoring the system, observing it at least once a year to study the evolution of the large-scale magnetic field of the star. A comparison between the evolution of the magnetic field of similar spectral type stars, with and without HJs, is the way to understand whether the presence of HJ influences the magnetic field generation (possibly due to tidal interactions).

This star is reported to show activity enhancement modulated by the orbital period for some observing epochs (\cite{shk03,shk05,shk08}). To study possible enhancement due to magnetospheric SPI, we studied the coronal large-scale magnetic field and the stellar activity. 

The coronal magnetic field is calculated using extrapolation techniques applied to the surface magnetic maps we derived. The field resembles that of a strongly tilted dipole. 

The stellar activity proxies (\caii and \halpha) show residual emission modulated by the rotation of the star to the first order. We looked for lower amplitude fluctuations by subtracting the rotational modulation to the residual emission. The new \caii and \halpha model residuals are periodic and correlated. The variation of these \halpha model residuals as a function of those of the \caii shows a trend different from the variation of the \halpha residual emission as a function of that of the \caii (having a higher slope value). The periods that best modulate these model residuals roughly match the beat period of the system, suggesting that a SPI magnetospheric interaction may be at the origin of such an effect. To confirm a SPI activity enhancement (or not), a larger data set is needed. 

 

Further observations of the large scale stellar magnetic field are required to study the magnetic cycle length, and possible influences of the planet on the stellar magnetic cycle. Combining observations at different wavelengths will allow us a detailed study of SPI, since it provides information for the system from the stellar surface to the corona.

\section*{Acknowledgments}
This work is based on observations obtained with ESPaDOnS at the Canada-France-Hawaii Telescope (CFHT). CFHT/ESPaDOnS are operated by the National Research Council of Canada, the Institut National des Sciences de l'Univers of the Centre National de la Recherche Scientifique (INSU/CNRS) of France, and the University of Hawaii. We thank the CFHT staff for their help during the observations. We also thank an anonymous referee for their comments on our manuscript. 
\bibliography{RFares_version5}
\bibliographystyle{mn2e}

\appendix
\section{Reconstructed map using lower degrees of spherical harmonics}
\label{lmax4}
We have reconstructed the magnetic map of HD~179949 using degrees of spherical harmonics up to an order smaller than the one used in the paper, for the purpose of comparison. Indeed,  when using $\ell_{max}=6$, only a small amount of energy is attributed to orders of $\ell > 4$ (about 7\% of the total energy). As shown in the formula used to derive $\ell_{max}$, it is always possible to resolve up to at least $\ell=4$, we therefor tested reconstructing the map with $\ell_{max}=4$ for both observing epochs. 

The reconstructed map is shown in Fig. \ref{Maplmax4}. Table \ref{tab:field_lmax4} lists the field properties using  $\ell_{max}=4$ and $\ell_{max}=6$ for comparison.

\begin{figure*}
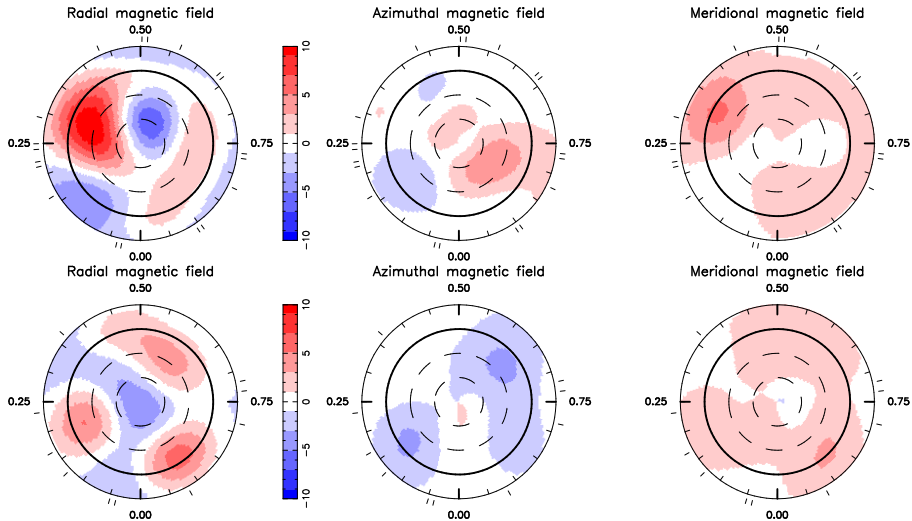

\center{\includegraphics[scale=0.5]{Maphd179949_sep09_lmax4.ps}
	\includegraphics[scale=0.5]{Maphd179949_jun07_lmax4.ps}}
\caption[]{The magnetic map of HD~179949 for 2009~September (upper panel) and 2007~June (lower panel), reconstructed using $\ell_{max}=4$. The reader can compare to Fig. \ref{fig:maps} where the maps were reconstructed using a higher order of spherical harmonics ($\ell_{max}=6$).}
\label{Maplmax4}
\end{figure*}

\begin{table*}
\caption[]{The field properties for the two epochs of observations for the reconstructed maps with $\ell_{max}=4$~and 6. The magnetic field strength averaged over the stellar surface (the error bar is of $\pm 0.3$~Gauss for both seasons) and the percentage of the poloidal energy relative to the total field energy are listed.  }
\begin{tabular}{ccccc}
&\multicolumn{2}{c}{$\ell_{max}=6$} & \multicolumn{2}{c}{$\ell_{max}=4$} \\
\hline
Epoch& B& $\rm E_{poloidal}$ &  B  & $\rm E_{poloidal}$ \\
& (G) & \% of $\rm  E_{total}$& (G) & \% of $\rm  E_{total}$\\
\hline
September~2009 & 3.7 & 90  &  3.9 & 90\\
\hline
June~2007 & 2.6 & 80  & 2.9 &83 \\
\hline
\end{tabular}
\label{tab:field_lmax4}
\end{table*}

We used the same method as in \ref{sec:DR} to quantify the DR parameters for 2009~September when using $\ell_{max}=4$. We find DR parameters of $\omeq=0.831\pm0.008~\rpd$ and $\dom=0.222 \pm 0.037~\rpd$, implying an equatorial rotation period of $7.56 \pm 0.07$~d and a polar rotation period of $10.3 \pm 0.8$~d, compatible with the ones found using $\ell_{max}=6$. These results are not surprising, since the reconstruction procedure uses Maximum-Entropy image reconstruction, and thus the simplest image compatible with the data is the one to be reconstructed. In our case, low orders are sufficient to describe the field in a reliable way. 

\section{Effect of low quality data on the field reconstruction}
\label{testZDI}
In section \ref{sec:maps}, we discuss briefly the effect of low quality data on the map reconstruction. Such effects were discussed elsewhere \citep{donati97b}, for an older version of ZDI code. In this appendix, we study the effect of low quality data (low S/N, sparse phase sampling), and show that the new version of the code (based on spherical harmonics decomposition) show similar trends as the old one when dealing with low quality data.

Two major cases will be studied, the first one is the effect of rotational phase gaps in observation on the reliability of the reconstructed map, and the second one is the reliability of a map reconstructed using a low S/N data set.

\subsection{Rotational phase gaps}
To test the effect of rotational phase gaps, we decide to use data from 2009~September data set, selecting spectra with good S/N ratio. We reconstruct two maps out of two sets of data, one with 9 spectra covering rotational phases between 0.2 and 0.8, and one with 7 spectra covering only half of the rotational phases (see Figure \ref{Fig:mapphase}). Compared to the map reconstructed from 19 spectra (section \ref{sec:maps}), both of these maps show a good recovering of the magnetic field of the observed part of the star. Even when half of the rotational cycle is not observed, the reconstructed field on the observed surface is very similar to the one for the map reconstructed with 19 spectra, but features of the unobserved surface are missed. 

\begin{figure*}
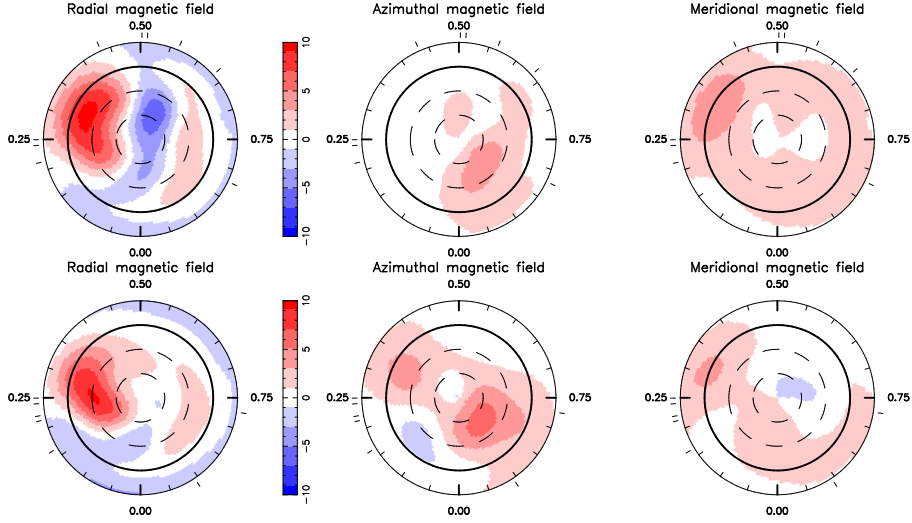

\center{\includegraphics[scale=0.5]{Map_test1_sep09.ps}
	\includegraphics[scale=0.5]{Map_test2_sep09.ps}}
\caption[]{The magnetic map of HD~179949 for 2009~September using 9 spectra with good S/N covering rotational phases from 0.2 to 0.8 (upper panel) and the reconstructed map for 2009~September using only 7 spectra with good S/N covering half of the rotation of the star. The features on the observed surface are well recovered compared to the map in \ref{sec:maps}.}
\label{Fig:mapphase}
\end{figure*}

\subsection{Low S/N}

One can also question how reliable is a reconstructed map based of spectra with low S/N ratio. In order to test this, we divide the 19 spectra data set of 2009~September to two completely independent subsets, one of 10 spectra with low S/N ratio, and a second one with 9 spectra with the highest S/N ratio of the original dataset. Table \ref{tab:SN} lists the properties of the field in each reconstructed map, i.e. magnetic energy and the fraction of the poloidal field. We reconstruct the magnetic maps using both subsets, using $\ell_{max}=6$~and $\ell_{max}=4$. 

\begin{table*}
\begin{tabular}{ccccc}
&\multicolumn{2}{c}{$\ell_{max}=6$} & \multicolumn{2}{c}{$\ell_{max}=4$} \\
\hline
Number of spectra & B& $\rm E_{poloidal}$ &  B  & $\rm E_{poloidal}$ \\
& (G) & \% of $\rm  E_{total}$& (G) & \% of $\rm  E_{total}$\\
\hline
19 (whole dataset) & 3.7 & 90  &  3.9 & 90\\
\hline
9 spectra with high S/N ratio & 3.7 & 90  & 3.9 &90 \\
\hline
10 spectra with low S/N ratio & 3.2 & 87 & 3.3  & 87 \\
\end{tabular}
\label{tab:SN}
\caption{The field characteristics when reconstructed using the whole dataset for 2009~September, a subset of only 9 spectra with the higher S/N ratio out of the 19 original spectra, and finally a subset of 10 spectra out of the 19 with the lower S/N ratio. The columns lists the mean magnetic energy and the energy in the poloidal component of the field, for the reconstructed maps using $\ell_{max}=6$~and $\ell_{max}=4$~respectively (see text for more details). }
\end{table*}

The magnetic energy reconstructed when using a low S/N dataset is, as clear from Table \ref{tab:SN}, lower than the energy reconstructed when using the whole dataset. However, the reconstructed maps (not shown here) recover most of the reconstructed field of the original map, showing the robustness of the reconstruction. These results are indeed similar to the one found by \cite{donati97b}.
\end{document}